\documentclass{aastex631}

\shorttitle{A radiatively driven UFO from RE\,J1034+396}
\shortauthors{Taylor et al.}

\begin{document}

\title{An Exceptionally Powerful, Radiatively Driven Ultrafast Outflow in the Rapidly Accreting AGN RE\,J1034+396}

\correspondingauthor{Chloe S. Taylor}
\email{ctaylor9@stanford.edu}

\author[0000-0003-3117-3476]{Chloe S. Taylor}
\affiliation{Department of Physics, Stanford University, 382 Via Pueblo Mall, Stanford, CA 94305, USA}
\affiliation{Kavli Institute for Particle Astrophysics \& Cosmology, Stanford University, 452 Lomita Mall, Stanford, CA 94305, USA}

\author[0000-0002-4794-5998]{Daniel R. Wilkins}
\affiliation{Kavli Institute for Particle Astrophysics \& Cosmology, Stanford University, 452 Lomita Mall, Stanford, CA 94305, USA}
\email{dan.wilkins@stanford.edu}

\author[0000-0003-0667-5941]{Steven W. Allen}
\affiliation{Department of Physics, Stanford University, 382 Via Pueblo Mall, Stanford, CA 94305, USA}
\affiliation{Kavli Institute for Particle Astrophysics \& Cosmology, Stanford University, 452 Lomita Mall, Stanford, CA 94305, USA}
\affiliation{SLAC National Accelerator Laboratory,  2575 Sand Hill Road, Menlo Park, CA 94025, USA}
\email{swa@stanford.edu}

\begin{abstract}

\noindent
We report the analysis of $\sim$1\,Ms of \textit{XMM-Newton} observations of the rapidly accreting active galactic nucleus RE\,J1034+396. The $0.3-9$\,keV EPIC-pn spectra are well described by a model consisting of steep continuum emission from the corona accompanied by relativistically-blurred reflection from a highly ionized accretion disk. The source is known to exhibit strong excess soft X-ray emission, which we show is well represented by thermal disk photons Comptonized by a warm plasma spanning the inner accretion flow.  Additionally, the EPIC-pn data provide compelling evidence ($\Delta C\sim60$ for 4 additional parameters) for the presence of an ultrafast outflow (UFO) with a line-of-sight velocity $v/c=0.307^{+0.001}_{-0.005}$, and an emission signature consistent with reflection of the corona from modestly ionized, outflowing gas. The simultaneous $0.5-2.5$\,keV RGS spectra show clear absorption lines.  Modelling of these data confirms the presence of the UFO and constrains its equivalent hydrogen column density, log $N_\mathrm{H}/$(atom cm$^{-2}$) = $21.7^{+0.1}_{-0.2}$. The RGS data also reveal at least two warm absorber components with a modest outflow velocity ($1680^{+40}_{-50}$ km/s).  The measured properties and time evolution of the UFO in RE\,J1034+396 suggest that it is formed from collisionally ionized plasma, launched from the disk surface and accelerated by radiation pressure.  The high terminal velocity and substantial absorbing column density imply that the outflow carries sufficient momentum and energy to transform its environment, being capable of driving out essentially all dust and gas it interacts with along the line of sight, even if the AGN were initially surrounded by a Compton thick absorber.

\end{abstract}

\section{Introduction} \label{sec:intro}
\noindent Studies of active galactic nuclei (AGN) undergoing periods of rapid accretion, at close to their Eddington limits, are central to understanding supermassive black hole growth and its impact on the evolution of their host galaxies \citep{OUTrev1,OUTrev2,OUTrev3}. The accretion disks of such systems can be extremely luminous and power fast, even relativistic, winds \citep{UFOobs}. However, the physical processes responsible for launching and driving these winds remain unclear.

Ultrafast outflows (UFOs) are some of the most energetic outflows detected in AGN.  They are characterized by mildly relativistic velocities, $v \sim 0.1-0.3c$, usually inferred from the detection of blueshifted Fe-K absorption lines in the $6-9$\, keV band, and typically high ionization states \citep{UFO1,UFO2,UFO3}. It has been postulated that UFOs may exist in over 35\% of nearby AGN \citep{UFOsearch1,UFOfrac1}.  While there are clues to the mechanisms that power UFOs \citep{UFOdrive1,UFOdrive2}, much remains to be understood about what drives these outflows and their impact on their surroundings.

RE\,J1034+396 is a narrow-line Seyfert~1 galaxy (NLS1) at a redshift $z = 0.04307$ \citep{z}.  NLS1 galaxies are characterized by high mass accretion rates onto moderately sized supermassive black holes \citep{NLS1}.  The central black hole in RE\,J1034+396 has a mass of 1-4 million solar masses \citep{goodM} and is accreting at roughly its Eddington limit \citep{newLedd}.  RE\,J1034+396 is known for being one of the few AGN with a clear quasi-periodic oscillation (QPO) in its X-ray emission\,\citep{qpodisc,allQPO}.  Previous work has reported evidence for warm absorbers, long lived outflows at velocities $v \sim 100-2000$ km s$^{-1}$ based on the presence of absorption lines detected in soft X-rays \citep{wa4,wa1,wa2,wa3,waufo}.  It has also been suggested that these warm absorber features may correlate with the QPO phase \citep{wa4}. Tentative evidence for a UFO in the excess variance spectrum was reported by \cite{varufo}.  \citet{wa3} suggest the potential existence of a UFO due to observed absorption features above $\sim7$ keV.  A further, recent claim argues for the presence of a UFO in absorption in RE\,J1034+396 in all epochs with velocity $v \sim 0.15c$ \citep{waufo}.

Here, we report on the analysis of recent deep ($\sim$1\,Ms of clean data) observations of RE\,J1034+396 made with the \textit{XMM-Newton} X-ray observatory.  We analyze the European Photon Imaging Camera pn CCD (EPIC-pn) and Reflection Grating Spectrometer (RGS) spectra, finding clear evidence for a UFO, detected via both emission and absorption lines, with velocity $v \sim 0.31c$, substantially higher than previous claims.   Given the extended duration of the observations, we are also able to probe the short and long term variability of the UFO, shedding new light on the mechanism by which it is powered. Our analysis shows that the spectrum of RE\,J1034+396 can be explained by a simpler model than has been employed in some previous studies.  This model consists of a primary corona, highly ionized reflection from the accretion disk, a warm Comptonizing plasma that spans the accretion disk surface, and the UFO.  A companion paper explores the implications of our simplified model for the origins of the QPO in this source (Taylor et al., in preparation).

\section{Observations and Data Reduction}

\noindent
RE\,J1034+396 was observed with the \textit{XMM-Newton} satellite \citep{XMM} for four continuous periods of 90, 87, 91, and 93 ks between 2020 November 20 and 2020 December 5. Six more observations of 94, 89, 93, 92, 94, and 94 ks were taken between 2021 April 26 and 2021 May 31 (Table~\ref{tab:obs}). Here, we use the data collected by both the EPIC pn camera \citep{PN}, which was operated in small window mode, and the Reflection Grating Spectrometers (RGS) \citep{RGS}.  

The observations were reduced using the XMM \textsc{science analysis system} (SAS) v21.0.0.  Event lists for the EPIC pn camera were processed and filtered using the \textsc{epproc} task and the latest calibration files.  Source photons were extracted from a 35 arcsec region centered on the point source.  We used the \textsc{evselect} task to extract spectra and the \textsc{rmfgen} and \textsc{arfgen} tasks to generate the response matrix and ancillary response. Spectral analysis was performed using \textsc{xspec} v12.13.0 \citep{xspec}.

\begin{table}[htbp]
    \centering
    \begin{tabular}{cccc}
        \hline
        Obs ID & Date (UT) & Duration (ks) \\
        (1) & (2) & (3) \\
        \hline
        \hline
        0865010101 & 2020-11-20 & 90\\
        0865011001 & 2020-11-30 & 87\\
        0865011101 & 2020-12-04 & 91\\
        0865011201 & 2020-12-02 & 93\\
        0865011301 & 2021-04-24 & 94\\
        0865011401 & 2021-05-02 & 89\\
        0865011501 & 2021-05-08 & 93\\
        0865011601 & 2021-05-12 & 92\\
        0865011701 & 2021-05-16 & 94\\
        0865011801 & 2021-05-30 & 94\\
        \hline
    \end{tabular}
    \caption{Summary of the \textit{XMM-Newton} observations.  The columns list (1) observation ID, (2) start date of the observation, and (3) duration of the EPIC-pn observation (rounded down to the nearest ks). }
    \label{tab:obs}
\end{table}

Data from the two \textit{XMM-Newton Reflection Grating Spectrometers}, RGS1 and RGS2, were gathered simultaneously to the EPIC pn camera. These were reduced following standard procedures, using the \textsc{sas} task \textsc{rgsproc} to perform the required filtering and calibration, extract source and background spectra, and generate instrumental response matrices. The default source and background regions for a point source located at the aim point were used. Additional filtering was performed to exclude time intervals during which the background count rate was greater than 0.2\,ct\,s$^{-1}$.

Due to damaged regions of each camera, there are gaps in the spectra collected by RGS1 and RGS2 at specific energies. Fortunately these differ for the two detectors. We therefore combine the spectra from RGS1 and RGS2 (from each of the spectral orders) using the \textsc{sas} task \textsc{rgscombine}, which sums the counts recorded in each spectral channel and generates an average response matrix for the two instruments. In addition to creating spectra for the 10 individual RGS observations, we also created a combined spectrum using the SAS tool \textsc{rgscombine}. Before analysis, the spectra were binned according to the optimal binning algorithm for sampling of the RGS instrument resolution\,\cite{kaastra_bleeker} using the \textsc{ftool} \textsc{ftgrouppha}. In this analysis, we specifically focus on the first order spectrum, again using \textsc{xspec} v12.13.0 \citep{xspec}.

\section{Spectral Analysis}
\subsection{The Background Spectrum}
\noindent
While RE\,J1034+396 is a bright source, the steep spectrum means that the source flux at high energies ($>7$ keV) is low.  This region of the spectrum is critical for detecting UFO features, so a rigorous treatment of the cosmic ray induced instrumental background proved necessary, including attention to possible contamination by soft proton flaring.

\subsubsection{Modelling the EPIC pn Particle Background}
\noindent
Above 8 keV, the source and particle background flux in the EPIC-pn data are comparable.  In order to perform an accurate analysis, we fit a full forward model for both the source and background components to the total observed spectrum, rather than subtracting a background spectrum.  Strong Cu K$\alpha$ and Zn K$\alpha$ instrumental lines appear in the EPIC-pn at 8.05 keV and 8.64 keV, respectively.  In order to quantify the instrumental background, including the Cu K$\alpha$ and Zn K$\alpha$ lines, we analyze 138 ks of small window mode, filter wheel closed data gathered across five observations between 2002 and 2021.  With the filter wheel set to the 'closed' position, source photons are blocked from reaching the XMM-Newton detector. Since the instrumental background varies across the detector, we extracted a particle background spectrum from the same 35 arcsec region on the detector that RE\,J1034+396 source photons were gathered from.  We modelled this background using a power law and two Gaussian lines corresponding to the Cu K$\alpha$ and Zn K$\alpha$ lines.  The normalization of the particle background is variable with time and depends on the solar cycle and solar activity on shorter timescales.  We therefore place a prior on the overall normalization of our model based on the solar cycle, using the ACIS High Energy Reject Rate from the \textit{Chandra X-ray Observatory} \citep{HERR}.

\subsubsection{Soft Proton Flaring in the EPIC pn data}
\noindent
In three of the 10 EPIC pn observations, at energies above 10 keV where the effective area of the \textit{XMM-Newton} mirrors is low, the observed count rate is more than double the expected particle background model count rate. Examination of the spectrum of this excess emission indicated that it is due to soft proton flares. To simplify our analysis, We dropped these compromised data from our study.  For the remaining seven observations, we restricted our analysis to energies between 0.3 and 9 keV, where the impact of uncertainties in our background modeling are negligible.  We note that all ten observations were used in the RGS analysis where the effects of soft proton flaring are not significant.

\subsection{Modeling the X-ray Spectrum}
\noindent
We first simultaneously fit the 0.3-9 keV EPIC pn spectra from the seven observations without significant soft proton flaring using a relativistic reflection model (\textsc{relxilllp}\footnote{We note that this model uses an emissivity profile assuming illumination of the disc by a point source.}) to describe the primary X-ray continuum from the corona and associated reflection from the inner accretion disk, including Doppler shifts and gravitational redshifts around the black hole \citep{REL1,REL2}. We also included a low temperature Comptonization component (\textsc{comptt}) for the significant soft excess emission observed, which is at a level significantly above that predicted by the reflection model alone.  Such a soft excess is thought to be generated through the repeated Compton scattering (or Comptonization) of thermal photons from the accretion disk by a warm plasma ($kT\sim0.2$ keV and $\tau \sim 15$) spanning the surface of the inner disk \citep{soft2}.  The \textsc{tbabs} model was used to account for absorption by the neutral phase of the Milky Way insterstellar medium along our line of sight, with the appropriate column density ($N_{\rm H} = 1.25 \times 10^{20}$ cm$^{-2}$; \citealp{hi4pi}).  We allow the \textsc{relxilllp} power law photon index, $\Gamma$, and normalization to vary between observations and tie the remaining relativistic reflection parameters.  We also let the Comptonizing plasma temperature, optical depth, and normalization vary between observations.  We tie the remaining parameters across all seven observations.  We note that in comparison to a simple power law model, use of the \textsc{relxilllp} to model to describe the primary continuum and associated reflection improves the C-statistic \citep{cstat} by 207 for 6 additional free parameters, demonstrating a clear statistical preference for this model. 

Having fitted this model, we observed a residual emission feature at $\sim$8.5 keV that we interpret as a blueshifted iron K$\alpha$ emission line from outflowing gas (for which the rest frame energy lies between 6.4 and 6.97\,keV depending on the ionization state). We modelled this feature using the \textsc{xillver} model for reflection from the surface of an ionized plasma \citep{XIL1,XIL2,XIL3}, blueshifted using the \textsc{zashift} model function in \textsc{xspec}. Tying all the parameters for the outflow across all seven observations, the addition of the \textsc{zashift*xillver} model improved the C-statistic by 57 for 4 additional free parameters.  Contributions to this improvement stem from both the $\sim$8.5 keV emission line region and blurred emission features at softer energies. We note that the (\textsc{zashift*xillver}) model provides a significantly better description of the emission signature of the outflow than a blueshifted Gaussian emission line (with the introduction of the latter only improving the C-statistic for the baseline model by 5 for 2 additional free parameters).

\begin{figure}[htbp]
\centering
\includegraphics[width=0.7\textwidth]{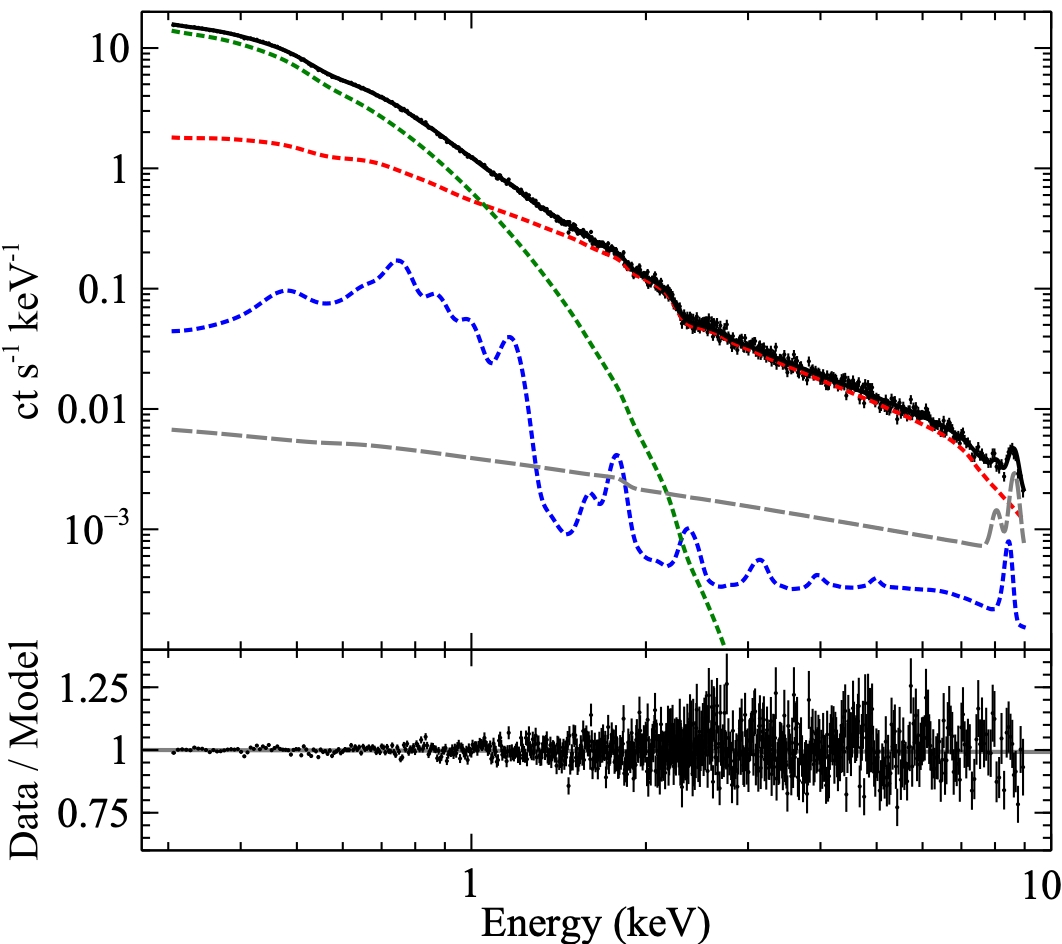}
\caption{The time averaged 0.3-9 keV EPIC pn X-ray spectrum of RE J 1034+396.  The dashed grey line shows the particle background model with instrumental lines.  The red dotted line shows the combined emission of the power law continuum and highly ionized reflection from the accretion disk.  The green dotted line shows the soft X-ray Comptonization component.  The blue dotted line shows the emission spectrum of the outflow.  
}
\label{fig:spm}
\end{figure}

Our final preferred model, \textsc{tbabs*(comptt+relxilllp+zashift*xillver)}, accounts for ISM absorption, the soft excess, the primary X-ray corona, relativistic reflection from the accretion disk, and blueshifted ionized reflection from outflowing gas (Fig.~\ref{fig:spm}a). The best fitting parameters for this model were found by minimizing the C-statistic \citep{cstat}.  Uncertainties were obtained using Markov Chain Monte Carlo (MCMC).  Markov chains were produced using the Goodman-Weare algorithm with 80 walkers, for 5000000 iterations after burning the first 5000 iterations.  The best-fitting model parameters and their uncertainties are shown in Table \ref{tab:param}.  The source is found to have relatively steep X-ray corona (time averaged photon index $\Gamma=2.15\pm0.02$). The ionization parameter associated with relativistically-blurred reflection from the accretion disk is quite high, log $\xi=3.52^{+0.01}_{-0.04}$, while the  iron abundance of the disk surface is $0.68\pm0.06$ solar (assuming the solar abundances from \cite{AFe}). The inclination angle, $i=51\pm1$ degrees, is tightly constrained. The input photon temperature, $kT_{0}=0.0474\pm 0.0003$\,keV, and time averaged Comptonizing plasma temperature, $kT_{\rm e}=0.204\pm 0.007$\,keV, and optical depth, $\tau=12.7^{+0.5}_{-0.3}$, of the soft \textsc{comptt} component are also well constrained by the data.  We note that the moderate optical depth means that small changes in the input photon temperature have significant effects on the shape of the spectrum, resulting in tight constraints on $kT_0$.  The outflowing gas detected in emission in the EPIC-pn data and modeled by the (\textsc{zashift*xillver}) components is shifted by $z = -0.272^{+0.001}_{-0.004}$, corresponding to a mean line-of-sight velocity, $v/c = 0.307^{+0.001}_{-0.005}$ after relativistic corrections, identifying the outflow as a UFO. Interestingly, the best fitting iron abundance of the outflow, $0.68\pm0.06$ Solar is consistent with that of the disk surface ($0.69\pm0.03$ Solar). 

\begin{table}[htbp]
    \centering
    \begin{tabular}{cccc}
        \hline
         \textbf{Model Component} & \textbf{Parameter} & \textbf{Tied Value} & \textbf{0865011001 Value}\\
         \hline
         \hline
         \textsc{TBABS} & Equivalent Hydrogen Column, $n_H$ ($10^{22} cm^{-2}$) & $0.025^*$\\
         \textit{(ISM Absorption)} & & \\
         \hline
         \textsc{COMPTT} & Input Photon Temperature, $kT_0$ (keV) & $0.0474^{+0.0003}_{-0.0003}$ &  \\
         \textit{(Soft X-ray Comptonization)} & Plasma Temperature, $kT_e$ (keV) & & $0.195^{+0.003}_{-0.002}$ \\
          & Plasma Optical Depth, $\tau_e$ & & $13.4^{+0.1}_{-0.2}$ \\
         \hline
         \textsc{RELXILLLP} & Corona Height, $h$ $(r_g)$ & --- \\
         \textit{(Power Law Emission}& Black Hole Spin, $a$ $(\frac{GM}{c^2})$ & --- \\
         \textit{and Disk Reflection)}& Inclination, $i$  (\text{deg}) & $51^{+1}_{-1}$ \\
         & Photon Index, $\Gamma$ & & $2.15^{+0.01}_{-0.01}$ \\
         & Ionization, log $\xi$ (erg cm $\text{s}^{-1}$) & $3.52^{+0.01}_{-0.04}$ \\
         & Iron Abundance, $A_{\text{Fe}}$ (solar) & $0.68^{+0.06}_{-0.06}$ \\
         & Reflection Fraction, $R$ & --- \\
         \hline
         \textsc{ZASHIFT} & Outflow velocity, $z$ & $0.272^{+0.001}_{-0.004}$\\
         \textit{(Outflow Blueshift)} & & \\
         \hline
         \textsc{XILLVER} & Iron Abundance, $A_{\text{Fe}}$ (solar) & $0.69^{+0.02}_{-0.03}$ \\
        \textit{(Ionized Reflection)} & Ionization, log $\xi$ (erg cm $\text{s}^{-1}$) & $1.22^{+0.02}_{-0.02}$ \\
         \hline
    \end{tabular}
    \caption{Best fitting model parameter values and 68.3\% confidence limits.  Parameters marked an asterisk are fixed at the stated value.  Parameters values marked with a dash are poorly constrained.  The last column shows sample values for the untied parameters from the 0865011001 observation.}
    \label{tab:param}
\end{table}

\subsection{Time Resolved Spectroscopy}
\noindent
We next applied our preferred model to the 0.3-9 keV EPIC-pn spectra from each of the seven individual observations of RE\,J1034+396 to determine the stability of the outflow velocity.  To do this, we used the fit described in the previous section, but untied the velocity of the outflow parametrized by the \textsc{zashift} component and the normalizaton of \textsc{xillver}.  The best fitting model parameters were again found by minimizing the C-statistic.  Uncertainties were obtained using MCMC.  The emission feature is detected significantly, improving the C-statistic by $>$10 for only 2 additional free parameters in all seven observations.

\begin{figure}[htbp]
\centering
\includegraphics[width=0.7\textwidth]{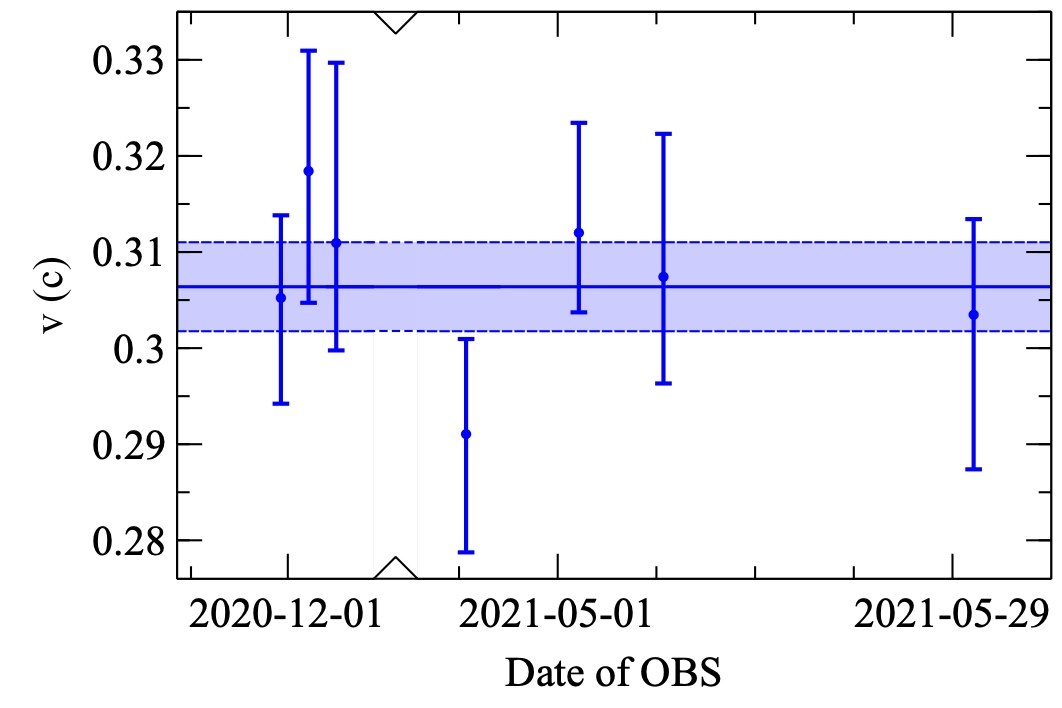}
\caption{Line-of-sight outflow velocity measurements from each EPIC-pn observation of RE\,J1034+396 as a function of time of observation.  The notches in the time axis indicate a gap in the axis between 2020 December 7 and 2021 April 21.  Velocities from emission features in the 0.3-9 keV EPIC pn spectra are in blue.  The average UFO velocities from the emission measurements are shown by the blue shaded bands. Error bars are 68.3\% confidence limits.}
\label{fig:ufo_z}
\end{figure}

To determine whether the outflow velocities measured independently across the seven EPIC pn observations were consistent, we performed a fit with a constant model, assuming Gaussian probability density functions (PDFs) for the outflow velocities.  The minimum $\chi^2$/dof = 3.24/6 = 0.54 obtained indicates that a constant model provides a statistically acceptable description of the data.  The average relativistically corrected velocity of the outflow across the seven observations is $v/c = 0.306\pm0.005$.  This is consistent with our measurement from the simultaneous modeling of the seven observations and suggests that the terminal velocity of the outflow is stable over at least month-long time scales (Fig.~\ref{fig:ufo_z}).  Several shorter, earlier \textit{XMM-Newton} observations of RE\,J1034+396 were made over the period 2009-2018. Applying the same analysis approach to these data shows that the terminal velocity of the outflow has remained consistent at $\sim0.31c$ for more than a decade.

\subsection{The RGS Spectrum}

\noindent 
Given the detection of a blueshifted emission feature from a UFO in the EPIC pn spectrum, we also analyzed the higher-resolution spectra provided by the RGS to search for additional spectral features that could be associated with the outflow.  We detected a variety of absorption features in the 0.6-1 keV range of the stacked RGS spectrum, where the instrument's effective area is largest.  We used the spectral synthesis code \textsc{cloudy} (ver. 23.01) to model this absorption, adopting the default solar abundances \citep{CLOUDY} \footnote{\textsc{cloudy} is designed to simulate the physical conditions within astronomical plasmas and predict their emission and absorption spectra \citep{cloudy_ori,cloudyOri}.}.  

We explored both photoionized and collisionally-ionized plasma models.  In the photoionized models, we constructed a model of the ionizing spectrum corresponding to the AGN soft excess, approximated by a black body at temperature $1.5\times10^6$K.  For simplicity, we did not include the X-ray power law component or reflection components of the AGN continuum in the \textsc{cloudy} modeling, since they do not significantly influence the absorption lines generated in the 0.5-1 keV energy range.  Photoionization equilibrium (PIE) is characterized by the ionization parameter, $\xi$, defined as

\begin{equation}
    \xi = \frac{L_{\mathrm{ion}}}{n_\mathrm{H}r^2},
\end{equation}

\noindent where $L_{\mathrm{ion}}$ is the 1-1000 Ry ionizing luminosity, $n_\mathrm{H}$ is the hydrogen number density at the cloud surface, and $r$ is the distance from the surface to the ionizing source \citep{xi,Lion}.  The hydrogen number density was fixed to $10^{10}$ cm$^{-3}$.  While the radiation field is responsible for ionizing the plasma in photoionized models, we do not necessarily expect the UFO plasma to be in thermal equilibrium with the radiation so we specify the kinetic temperature of the gas in our models.  We created a grid of models that were used in \textsc{xspec} to fit the observed spectrum, varying the ionization parameter $\xi$ [log $\xi$/(erg cm s$^{-1}$) = 0.5-2, $\Delta=0.25$], temperature T [log T/(K) = 6.0-6.5, $\Delta=0.1$], hydrogen column density $N_\mathrm{H}$ [log $N_\mathrm{H}$/(cm$^{-2}$) = 20-22, $\Delta=0.25$], and turbulent line width $\sigma$ [log $\sigma$/(km s$^{-1}$) = 1.5-3.5, $\Delta=0.5$].

The \textsc{cloudy} code also enables the exploration of collisional ionization equilibrium (CIE) models, for which the temperature of the gas fully determines its ionization state.\footnote{We note that the timescale for the emerging plasma to achieve CIE near the surface of the disk is rapid \citep{tCIE}.}  Previous work has found evidence of collisionally ionized warm absorbers in AGN produced through shocks or in high density gas \citep{collout1,collout2}.  We created a grid of models varying the temperature T [log T/(K) = 6.0-6.5, $\Delta=0.1$], hydrogen column density $N_\mathrm{H}$ [log $N_\mathrm{H}$/(cm$^{-2}$) = 20-22, $\Delta=0.25$], and turbulent line width $\sigma$ [log $\sigma$/(km s$^{-1}$) = 1.5-3.5, $\Delta=0.5$].

\begin{figure}[htbp]
\centering
\includegraphics[width=\textwidth]{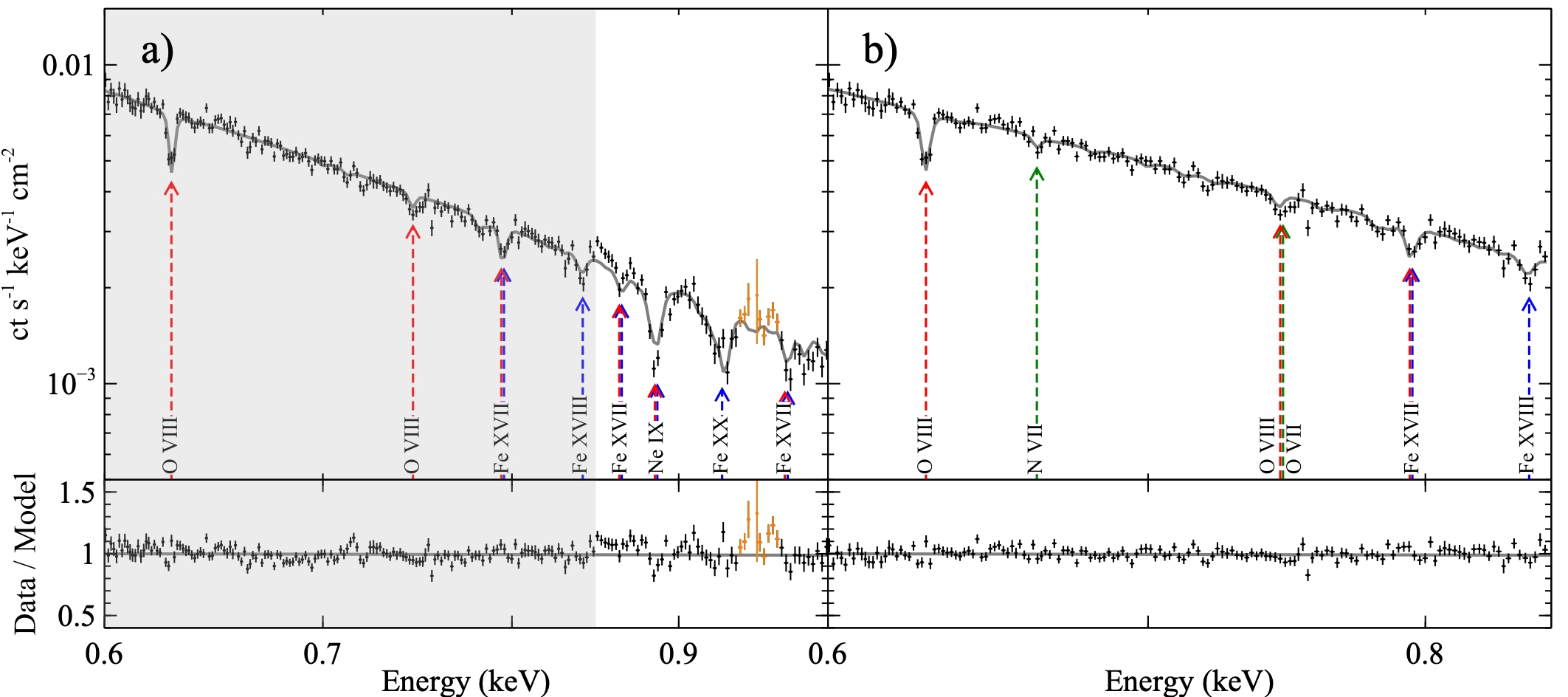}
\caption{a) The time averaged 0.6-1 keV X-ray spectrum of RE\,J 1034+396 from the RGS observations.  Absorption lines associated with the photoionized warm absorber are indicated by the blue dashed lines.  Absorption lines associated with the collisionally ionized warm absorber are indicated by the red dashed lines.  Orange data points denote instrumental features. b) The time averaged 0.6-0.85 keV X-ray spectrum of RE\,J 1034+396 from the RGS observations (covering the grey shaded region in the left hand panel).  Absorption lines associated with the photoionized warm absorber are indicated by the blue dashed lines.  Absorption lines associated with the collisionally ionized warm absorber are indicated by the red dashed lines.  Absorption lines associated with the $\sim0.31c$ UFO are indicated by the green dashed lines.}
\label{fig:rgs}
\end{figure}

For the RGS analysis, we used a simple power law to model the 0.6-1 keV continuum emission and the \textsc{tbabs} model to account for Galactic absorption by the neutral phase of the interstellar medium.  We note that the power law provides a reasonable approximation to this limited region of the spectrum. The 0.94-0.97 keV portion of the spectrum is excluded in our fits due to instrumental features in the effective area associated with chip gaps.  Adding a single photoionized warm absorber at the redshift of RE\,J1034+396 to the power law model reduces the modified Cash statistic by 438 for 5 degrees of freedom.  However, this model is unable to reproduce a strong O VIII absorption line observed around 600 eV.  Adding a second photoionized warm absorber reduces the modified Cash statistic by another 83, for 5 degrees of freedom.  However, this model is still unable to reproduce a strong O VIII absorption line around 600 eV.  Replacing both photoionized warm absorbers with a single collionally ionized warm absorber again improves on the fit over a simple power law model, reducing the modified Cash statistic by 349 for 4 degrees of freedom. However, this model is unable to reproduce a strong Ne IX line observed around 885 eV. In contrast, introducing a warm absorber with both photoionized and collisionally ionized components, where the velocity, temperature, and turbulent line width parameters are tied between components, provides a much better description of the absorption spectrum.  Compared to the power law plus \textsc{tbabs} model, the modified Cash statistic is reduced by 622 for 6 degrees of freedom (Fig.~\ref{fig:rgs}a).  The introduction of additional photoionized warm absorber components did not improve the fit significantly. 

We note that the temperature associated with these warm absorbers (likely at radii $\sim10^6 r_\mathrm{g}$), $kT = 0.24\pm0.02$ keV, is consistent with that of the \textsc{comptt} component that spans the inner disk surface ($\sim10-100 r_\mathrm{g}$). It is also consistent with the temperature inferred from optical and UV Fe XI lines emitted from the broad-line region ($\sim10^3-10^5 r_\mathrm{g}$) \citep{BLR}.  Additionally, the velocity of the warm absorbers, $v=1730^{+40}_{-50}$ km/s, is consistent with that inferred from the Fe XI lines \citep{BLR}.  Although we identify similar absorption lines at a consistent velocity to the warm absorbers discussed in \cite{waufo,wa3}, the best fitting ionization parameter of our photoionized warm absorber is significantly lower.  These previous warm absorber studies did not consider collisionally ionized warm absorbers which could explain the differences in the best fitting photoionized models.

In order to search for evidence of absorption lines associated with the UFO, we next concentrated our analysis on the cleanest region of the spectrum from 0.6-0.85 keV. In doing so, we imposed Gaussian priors on the PIE warm absorber component based on the full 0.6-1 keV fit (the most significant absorption features constraining this component occur above 0.85 keV). Refitting the data for this region and introducing an additional collisionally ionized component improves the modified Cash statistic by 11 for 4 degrees of freedom (Fig.~\ref{fig:rgs}b). 
 Due to the small improvement in the modified Cash statistic between models, we use the Akaike information criterion (AIC) to compare the relative quality of the models \citep{AIC}.  The AIC is defined as AIC = $2k-2\mathrm{ln}(L)$, where $k$ is the number of parameters in the model and $L$ is the maximized value of the Likelihood function for the model.  The relative likelihood of a model $i$ compared to a model $j$ is exp((AIC$_j$-AIC$_i$)/2).  Application of the Akaike information criterion (AIC) indicates a modest statistical preference (the model with the additional absorber reducing the AIC by $\sim3$ and being five times more likely).  Adding a further photoionized component did not provide a similar improvement to the fit.  Critically, however, the relativistically adjusted velocity of $0.3097^{+0.0011}_{-0.0005}c$ associated with this additional collisionally ionized absorber is in precise agreement with the average velocity of the UFO inferred from the seven independent 0.3-9 keV EPIC pn emission spectra, confirming that the same outflowing material is detected by both instruments.  We note that the velocity of this UFO is roughly twice the velocity of the UFO detected in absorption by \cite{waufo}, indicating that our detection is not related to the detection in previous work.

The preferred model for the UFO is characterized by a temperature log T/(K) = $6.70^{+0.05}_{-0.02}$ and a column density log $N_\mathrm{H}/$(cm$^{-2}$) = $21.71^{+0.12}_{-0.18}$.  We also infer an opacity of $\sim5-12\,\kappa_T$, where $\kappa_T$ is the Thomson opacity.  This enhancement in opacity above the Thomson opacity is due to bound-free radiatitive processes.  The best fitting CIE temperature of 0.43 keV for the UFO is slightly warmer than the temperature of the Comptonized emission component associated with the warm atmosphere spanning the surface of the disk \citep{soft,soft2}.  Together with the aforementioned iron abundance constraints, this paints a picture where the material in the wind is first lifted from the disk surface, possibly by magneto-hydrodynamic processes associated with a magnetically choked accretion flow \citep{MCAF}, and then accelerated by radiation pressure.  We conclude that our outflow is collisionally ionized and that the column density of the outflow is log $N_\mathrm{H}/$(cm$^{-2}$) $\approx$ 21.7.

\subsection{Acceleration of the outflow}\label{acceleration}

\begin{figure}[htbp]
\centering
\includegraphics[width=0.7\textwidth]{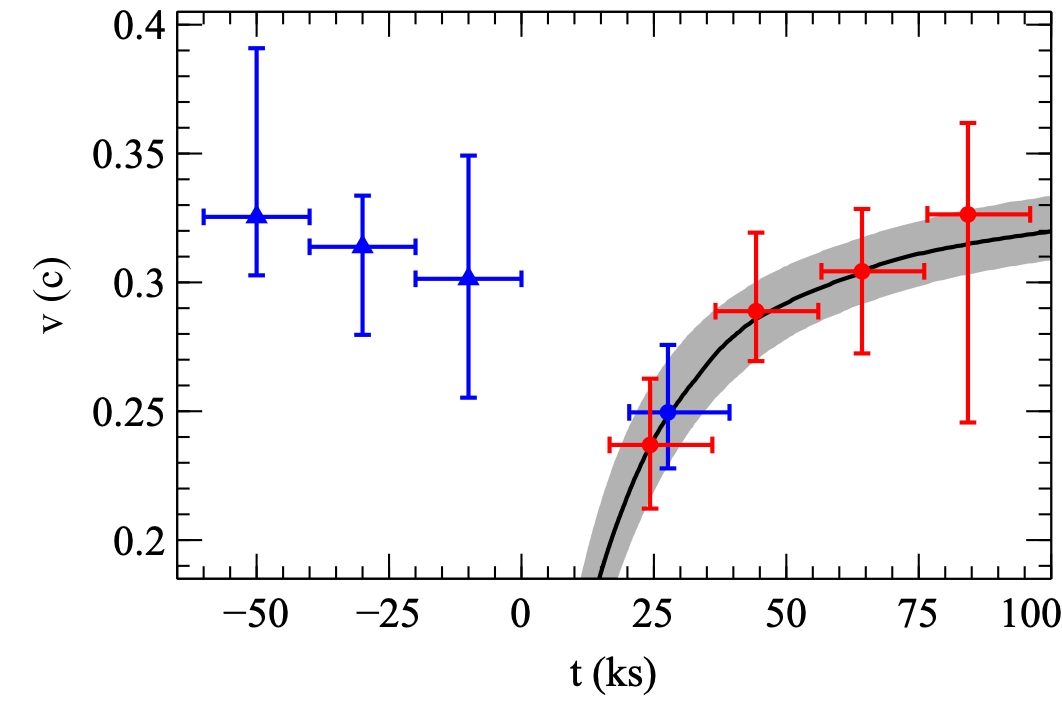}
\caption{Outflow velocity as a function of time for two separate EPIC pn observations.  An ejection event occurs at $t=0$.  The early blue triangular points (consistent with a constant velocity $\sim 0.31c$) show the outflow at its terminal velocity before the launch of new material.  The circular points show the outflow's velocity after a new launch of material.  The black line indicates the best fitting model, with 68.3\% uncertainties shown in grey.}
\label{fig:ufo_acc}
\end{figure}

\noindent In order to probe variability of the UFO on even shorter time scales, we next divided each of the seven $\sim$90\,ks EPIC pn observations into four $\sim20$\,ks segments, fitting each with our preferred model.  Within five of the observations, we find all the parameters, including the outflow velocity, to be stable.  The remaining two observations exhibit variability in the outflow velocity, with each containing a time segment where the outflow velocity is significantly slower than the inferred line of sight terminal velocity of $v/c \sim 0.31$ ($v/c = 0.24^{+0.03}_{-0.02}$ and $0.25^{+0.03}_{-0.02}$, respectively).  The time variability across these two observations (OBS IDs 0865011101 and 086511801) is plotted in Figure~\ref{fig:ufo_acc}.   If the observed variability is due to the episodic ejection of material from the disk surface followed by its acceleration to relativistic speeds, we would expect short time periods when the outflow is accelerating and longer periods when the outflow is at its terminal velocity.  Thus, on month-long timescales the velocity should appear stable, while shorter time segments reveal the acceleration.  We note that the QPO period of $\sim$4 ks is not obviously related to the wind launch timescale.

The two commonly proposed driving mechanisms for UFOs are radiation pressure and acceleration by magnetic fields \citep{acc,mag}.  Assuming a radiatively driven wind from an axisymmetric thin accretion disk, the outflow acceleration can be determined given five parameters: the mass and luminosity of the AGN, the outflow launch radius, the initial velocity of the outflow, and the opacity of the wind material (see Appendix~\ref{sec:acc_mod}).  RE\,J1034+396 has a bolometric luminosity of $\sim 5 \times 10^{44}$ erg s$^{-1}$ \citep{newLedd}.  The mass of the black hole in RE\,J1034+396 is estimated to be $1-4 \times 10^{6}$ \(M_\odot\) from measurements of the stellar bulge velocity dispersion and from the second moment of the H$\beta$ line \citep{goodM}.  For this calculation, we assume a maximal mass of $4 \times 10^{6}$ \(M_\odot\) and an accretion rate, L/L$_{Edd}$ $\sim$ 1, noting also that, given the high accretion rate, the thickness of the accretion disk in RE\,J1034+396 may be enhanced as radiation pressure within the disk becomes dominant \citep{disk1,disk2}.  Initially, we assign the outflow no radial velocity, and equal rotational velocity and velocity in the direction normal to the accretion disk surface,  $v_\varphi=v_z=\sqrt{r_G/r}$.  The relatively large initial value of $v_\varphi$ has minor impacts on the model, reducing the best fitting wind opacity and wind angle.  This leaves two remaining parameters to describe the outflow evolution, the launch radius and wind opacity.

The best fitting launch radius and wind opacity are determined by fitting the radiatively driven outflow model to the wind velocities from the three time variable observations.  We assume the wind velocity is entirely along our line of sight.  Given the measured disk inclination of $51 \pm 1$ deg (see Table~\ref{tab:param}), this is consistent with the best fitting wind angle of $64^{+3}_{-4}$ deg after accounting for a $\sim$10 deg pitch angle. For the blue data in Figure~\ref{fig:ufo_acc}, we ignore the three velocity measurements that occur before the drop in velocity.  These are likely to be associated with previously launched wind material already at its terminal velocity, before the new release of material from the accretion disk.  We perform a Markov chain Monte Carlo (MCMC) sampling of the parameter space, including the relative time offsets of the three data sets as free parameters.  As expected, the results exhibit a degeneracy between the the launch radius and wind opacity.  Fig.~\ref{fig:ufo_acc} shows the best fitting acceleration curve with 68.3\% confidence uncertainties for the preferred time offsets of $t_0 = 24^{+12}_{-8}$ ks and $28^{+12}_{-7}$ ks for the red and blue points respectively.  We can further constrain the launch radius and wind opacity by introducing a terminal velocity constraint: here, we use the average wind velocity from the 7 observations, and assume that this terminal velocity is reached by $t = 10^4$ ks.  This assumption holds because by $t = 10^4$ ks the velocity appears to be constant for all modeled acceleration curves.  With this constraint, the wind launch radius and opacity are $r = 112^{+39}_{-38}r_G$ and $\kappa = 11^{+5}_{-4}\kappa_T$, where $r_G$ is the gravitational radius and $\kappa_T$ is the Thomson opacity.  We note that the opacity is in good agreement with the 0.01-10 keV Rosseland mean opacity predicted by the \textsc{cloudy} modeling ($\sim5-12\,\kappa_T$), where the range comes from uncertainty in the hydrogen column density.

\section{Discussion}

\begin{figure}[htbp]
\centering
\includegraphics[width=0.7\textwidth]{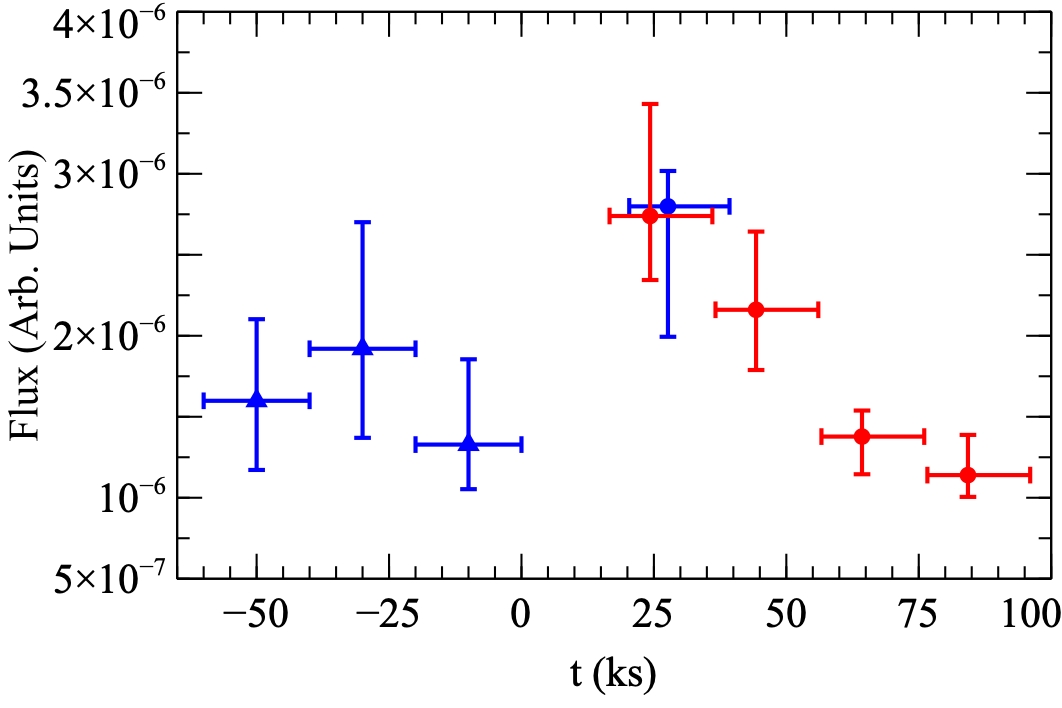}
\caption{The flux of the \textsc{xillver} component associated with reflection off of dense clumps in the outflow as a function of time for the two separate EPIC pn observations.  An ejection event occurs at $t=0$.  The early blue triangular points show the low flux associated with previously accelerated material before the new launch.  The circular points show the flux evolution associated with newly launched material.}
\label{fig:flux}
\end{figure}

\noindent
While a single phase, collisionally ionized plasma model with an outflow velocity of $\sim 0.31 c$ and a temperature of $\sim 5\times 10^6 K$, slightly hotter than at the accretion disk surface, is able to describe the RGS absorption features associated with the UFO in RE\,J1034+396 impressively well, in detail the outflow will be clumpy. Due to the high ionizing flux, reflection from the surface of clouds, or clumps, in the outflow located close to the AGN will generate the emission signatures of the outflow observed in the EPIC-pn spectra and modelled with the \textsc{xillver} component.  Fig.~\ref{fig:flux} shows how the flux associated with the \textsc{xillver} component peaks and then decreases for the ejection events shown in Fig.~\ref{fig:ufo_acc}.  This is expected because the clumps will spread out as material accelerates away, resulting in less reflection, meaning less flux from the \textsc{xillver} component.  The relatively strong observed emission features mean that these denser clumps in the outflow can be expected to have a modest covering fraction.

The absorption lines observed in the RGS data are expected to be generated predominantly by the more diffuse gas in the outflow, which may have a substantial covering fraction. The absence of evidence for temperature variability in the absorbing material over the duration of the observations \footnote{We observe signs of the N VII and O VII lines associated with the UFO consistently across the individual observations with no other strong lines appearing, indicating no significant temperature variability in the absorbing material.} suggests that the cooling time for the diffuse outflowing gas is relatively long, likely of order at least a year. In this case, the observed temperature and column density of the outflow would suggest a typical density for the diffuse outflowing gas of order $n_{\rm e} \sim 10^5$\,cm$^{-3}$.  This ensures that the cooling time exceeds the time to CIE \citep{tCIE}, consistent with the observed lack of temperature variability.

\begin{figure}[htbp]
\centering
\includegraphics[width=0.7\textwidth]{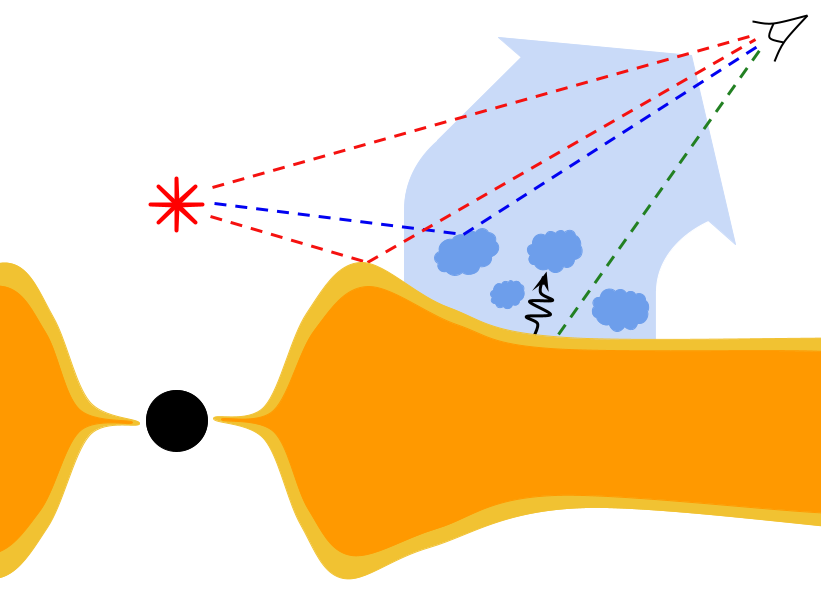}
\caption{Schematic of the inner regions of RE\,J1034+396. X-rays emitted by the corona are observed both directly and reflected off the surface of the highly ionized inner accretion disk (red dashed lines).  We do not intend to make any strong inferences about the coronal geometry, but simply illustrate the corona to be representative of a relatively compact central X-ray source.  Soft excess X-ray emission is produced by the Comptonization of disk photons in a warm $\sim 0.2$ keV plasma that spans the surface of the disk. Emission features associated with the UFO are produced via ionized reflection from the skin of the nearest, densest clumps of outflowing material (blue dashed lines). Absorption lines produced in the bulk of the UFO are seen at soft energies (green dashed line).  Once launched, the outflow is accelerated by radiation pressure from the disk.  We note that the warm absorber may emerge from the disk in a similar manner, but at much larger radii (not drawn here).  The thermal disk is shown in orange and the surface Comptonizing plasma is shown in gold.}
\label{fig:dia}
\end{figure}

Fig.~\ref{fig:dia} shows schematically where each of the spectral features originates.  The harder X-ray emission at energies $E>2$\,keV is dominated by direct coronal emission and reflection from the highly ionized disk surface (red dashed), modelled by the \textsc{relxilllp} component. Unlike many AGN, distinct reflection features from the surface of the accretion disk are largely absent\footnote{The broad Fe K line and Compton hump typically associated with disk reflection are also absent in the 3-50 keV \textit{NuSTAR} spectrum.}; this is due to the highly ionized state of the inner disk surface, which in turn stems from the high accretion rate and associated X-ray luminosity. We note, however, that the blue wing of the Fe K reflection region remains sufficiently distinct to place a relatively tight constraint on the disk inclination. The soft excess, modelled by the \textsc{comptt} component, that dominates the emission below $1$\,keV is caused by the Comptonization of thermal disk photons by the warm $\sim 0.2$\,keV plasma spanning the disk surface. The absorption lines in the RGS spectrum are generated as X-ray emission from the source passes through the diffuse outflow (green dashed). The emission features from the outflow are produced by X-rays from the hot corona reflecting off the denser clumps of gas within the UFO in the closest regions to the AGN (blue dashed).

The ionization state of the wind in RE\,J1034+396 is lower than reported for most other UFOs, identified on the basis of blueshifted Fe XXV and XXVI K-shell absorption lines. The more temperate ionization state and the strong soft X-ray emission from the surface of the disk in RE\,J1034+396 enables efficient radiative acceleration of the wind up to relativistic velocities \citep{windsim}. We anticipate that the UFO plasma will be in CIE as it is lifted up from the disk.  As it is bombarded by the radiation field, it will depart from its CIE state and eventually reach PIE.  

 
It is also interesting to consider the ability of the UFO in RE\,J1034+396 to impact its host galaxy.  Assuming a typical final wind speed for material escaping the host galaxy of $\sim$1000 km/s, a momentum conserving ($N_\mathrm{UFO} v_\mathrm{UFO} = N_\mathrm{gal} v_\mathrm{gal}$) UFO in RE\,J1034+396 could accelerate column densities of $\sim$$10^{24}$ atom cm$^{-2}$. For an energy conserving ($N_\mathrm{UFO} v^2_\mathrm{UFO} = N_\mathrm{gal} v^2_\mathrm{gal}$) outflow, the UFO could accelerate even larger column densities, approaching $\sim$$10^{26}$ atom cm$^{-2}$. Thus, along any line of sight that the UFO operates, we have a mechanism that is more than capable of removing sufficient gas to effectively stifle star formation and limit the fuel supply available for future episodes of rapid supermassive black hole accretion growth.

\section{Conclusions}

\noindent 
We have reported the discovery of a UFO with a line of sight velocity of $\sim 0.31c$ in \textit{XMM-Newton} EPIC-pn and RGS observations of the nearby, rapidly accreting narrow-line AGN, RE\,J1034+396. The outflow is detected consistently, and with high significance, in both emission and absorption. The discovery of a UFO in a nearby, rapidly accreting AGN provides fresh insights into feedback mechanisms. The short timescale acceleration seen in the line-of-sight outflow velocity is consistent with that of a radiatively driven wind. The high opacity of this wind, inferred from the line-rich RGS spectrum, and strong soft X-ray luminosity both help to facilitate efficient radiative acceleration \citep{radsim}.  The long term stability of the outflow's terminal velocity over a period of years also supports the radiation driving picture. However, contributions from magnetic forces may also be essential in initially lifting the UFO material from the disk surface, where it can be accelerated by radiation pressure.

The proximity of RE\,J1034+396 provides a close-up view of processes expected to be widespread in the early Universe, when highly luminous AGN with rapidly growing black holes were more common. Our  absorption modeling suggests an equivalent hydrogen column density of $N_{\rm H} \sim 10^{22}$ atom cm$^{-2}$ associated with the outflow.  Given this high velocity and substantial column density,  the UFO in RE\,J1034+396 should be able to clear out substantial column densities of material ($N_{\rm H} \gtrapprox$$10^{24}$ atom cm$^{-2}$) surrounding the central source. The ability of sources like RE\,J1034+396,
powered by near-Eddington or super Eddington accretion to effectively clear their environments is consistent with previous work that has found very few highly obscured, high Eddington ratio sources in the low-$z$ \textit{Swift} BASS AGN survey \citep{BASS}. Our results underscore the importance of periods of high accretion in feeding supermassive black hole growth and shaping the environments we see today.

\begin{acknowledgments}
We thank Zhefu Yu, Adam Mantz, and Roger Blandford for helpful discussions. This work was supported by the NASA Astrophysics Data Analysis Program under grant number 80NSSC22K0406. CST is supported by the ABB Stanford Graduate Fellowship.  SWA is supported in part by the U.S. Department of Energy under contract number DE-AC02-76SF00515.
\end{acknowledgments}

\vspace{5mm}
\facility{XMM-Newton}


\software{XSPEC \citep{xspec}, Cloudy \citep{CLOUDY}}

\appendix

\section{Outflow Acceleration Model}
\label{sec:acc_mod}

\noindent
We model the acceleration of a UFO driven by radiation pressure from the accretion disk.  Radiatively driven outflows are expected to reach a terminal velocity as their acceleration decreases over time.  As the outflow speeds up, disk radiation is redshifted to lower energy, flux, and photon momentum flux in the outflow's rest frame.  These effects, in addition to the increasing distance from the radiation source, mean that the wind feels a smaller radiation pressure over time resulting in a decrease in acceleration.  Following \cite{acc}, we obtain velocity curves by numerically solving the relevant differential equations.  We assume an axisymmetric wind launched from a geometrically thin accretion disk and use cylindrical coordinates $(r,\phi,z)$.  Using Euler's equation for an ideal fluid, we can write the equation of motion for the wind as

\begin{equation}
    \frac{\textrm{d}v}{\textrm{d}t} = \frac{L'\kappa}{4\pi (r^2+z^2) c} - \frac{GM}{r^2+z^2},
\end{equation}

\noindent where $L'$ is the AGN luminosity in the frame of the wind and $\kappa$ is the wind opacity.  Accounting for the effects of special relativity, $L'=\frac{L}{\gamma^4(1+\beta cos(\theta))^4}$, where $L$ is the luminosity of the AGN in the source frame, $\gamma$ is the Lorentz factor, $\beta = \frac{v}{c}$, and $\theta$ is the angle between the velocity of the gas and the incident luminosity.  We rescale the opacity to be in units of the Thomson opacity, $\kappa' = \frac{\kappa}{\kappa_T}=\kappa\frac{m_p}{\sigma_T}$, and the luminosity to be in terms of the Eddington ratio, $\lambda_{\mathrm{Edd}}=\frac{4\pi GMm_pc}{\sigma_T}$.  This results in the following set of differential equations: 

\begin{equation}
    \frac{\textrm{d}v_r}{\textrm{d}t} = \left(\frac{\lambda_\mathrm{Edd}\kappa'}{\gamma^4(1+\beta \cos(\theta))^4}-1\right)\frac{r}{(r^2+z^2)^{3/2}}+\frac{l^2}{r^3} ,
\end{equation}

\begin{equation}
    \frac{\textrm{d}v_z}{\textrm{d}t} = \left(\frac{\lambda_\mathrm{Edd}\kappa'}{\gamma^4(1+\beta \cos(\theta))^4}-1\right)\frac{z}{(r^2+z^2)^{3/2}},
\end{equation}

\noindent where $l$ is the specific angular momentum, a conserved quantity.  In these equations, $r$ and $z$ are in units of $r_G=\frac{GM}{c^2}$ and $t$ is in units of $t_G=\frac{r_G}{c}$.  We impose the initial conditions $(r_0,0,0)$ and $(0,v_{\varphi},v_{z,0})$ where $v_{\varphi} = v_{z,0} = \sqrt{\frac{1}{r_0}}$, $r_0$ is in units of $r_G$ and velocities are in units of $c$.  For a given supermassive black hole mass and luminosity, wind opacity and launch radius, we can numerically solve these equations.  For our MCMC sampling, we integrate the equations for 100ks in $10^4$ logarithmically spaced time steps.  This resolution is chosen to maximize efficiency without noticeably altering the solution.  For the MCMC sampling with the additional terminal velocity constraint, we integrate the equations for $10^4$ks in $2.5\times10^4$ logarithmically spaced time steps.

In our analysis, we assume a maximal mass of $4 \times 10^{6}$ \(M_\odot\) and an accretion rate, L/L$_{Edd}$ $\sim$ 1 to determine the preferred wind opacity and launch radius.  If we instead assume a minimal mass off $10^{6}$ \(M_\odot\), we obtain an equivalently good fit but with an increased preferred launch radius and wind opacity, again with large degeneracy.  While it is difficult to constrain the launch radius and wind opacity in the absence of a precise black hole mass constraint, the radiation driven wind acceleration curves are clearly good fits to the data and suggest that the UFO is likely being driven by radiation pressure from the disk.

\bibliography{ref}{}

\begin{thebibliography}{}
\expandafter\ifx\csname natexlab\endcsname\relax\def\natexlab#1{#1}\fi
\providecommand{\url}[1]{\href{#1}{#1}}
\providecommand{\dodoi}[1]{doi:~\href{http://doi.org/#1}{\nolinkurl{#1}}}
\providecommand{\doeprint}[1]{\href{http://ascl.net/#1}{\nolinkurl{http://ascl.net/#1}}}
\providecommand{\doarXiv}[1]{\href{https://arxiv.org/abs/#1}{\nolinkurl{https://arxiv.org/abs/#1}}}

\bibitem[{{Ahn} {et~al.}(2012){Ahn}, {Alexandroff}, {Allende Prieto}, {Anderson}, {Anderton}, {Andrews}, {Aubourg}, {Bailey}, {Balbinot}, {Barnes}, {Bautista}, {Beers}, {Beifiori}, {Berlind}, {Bhardwaj}, {Bizyaev}, {Blake}, {Blanton}, {Blomqvist}, {Bochanski}, {Bolton}, {Borde}, {Bovy}, {Brandt}, {Brinkmann}, {Brown}, {Brownstein}, {Bundy}, {Busca}, {Carithers}, {Carnero}, {Carr}, {Casetti-Dinescu}, {Chen}, {Chiappini}, {Comparat}, {Connolly}, {Crepp}, {Cristiani}, {Croft}, {Cuesta}, {da Costa}, {Davenport}, {Dawson}, {de Putter}, {De Lee}, {Delubac}, {Dhital}, {Ealet}, {Ebelke}, {Edmondson}, {Eisenstein}, {Escoffier}, {Esposito}, {Evans}, {Fan}, {Femen{\'\i}a Castell{\'a}}, {Fern{\'a}ndez Alvar}, {Ferreira}, {Filiz Ak}, {Finley}, {Fleming}, {Font-Ribera}, {Frinchaboy}, {Garc{\'\i}a-Hern{\'a}ndez}, {Garc{\'\i}a P{\'e}rez}, {Ge}, {G{\'e}nova-Santos}, {Gillespie}, {Girardi}, {Gonz{\'a}lez Hern{\'a}ndez}, {Grebel}, {Gunn}, {Guo}, {Haggard}, {Hamilton}, {Harris}, {Hawley}, {Hearty}, {Ho}, {Hogg}, {Holtzman},
  {Honscheid}, {Huehnerhoff}, {Ivans}, {Ivezi{\'c}}, {Jacobson}, {Jiang}, {Johansson}, {Johnson}, {Kauffmann}, {Kirkby}, {Kirkpatrick}, {Klaene}, {Knapp}, {Kneib}, {Le Goff}, {Leauthaud}, {Lee}, {Lee}, {Long}, {Loomis}, {Lucatello}, {Lundgren}, {Lupton}, {Ma}, {Ma}, {MacDonald}, {Mack}, {Mahadevan}, {Maia}, {Majewski}, {Makler}, {Malanushenko}, {Malanushenko}, {Manchado}, {Mandelbaum}, {Manera}, {Maraston}, {Margala}, {Martell}, {McBride}, {McGreer}, {McMahon}, {M{\'e}nard}, {Meszaros}, {Miralda-Escud{\'e}}, {Montero-Dorta}, {Montesano}, {Morrison}, {Muna}, {Munn}, {Murayama}, {Myers}, {Neto}, {Nguyen}, {Nichol}, {Nidever}, {Noterdaeme}, {Nuza}, {Ogando}, {Olmstead}, {Oravetz}, {Owen}, {Padmanabhan}, {Palanque-Delabrouille}, {Pan}, {Parejko}, {Parihar}, {P{\^a}ris}, {Pattarakijwanich}, {Pepper}, {Percival}, {P{\'e}rez-Fournon}, {P{\'e}rez-R{\`a}fols}, {Petitjean}, {Pforr}, {Pieri}, {Pinsonneault}, {Porto de Mello}, {Prada}, {Price-Whelan}, {Raddick}, {Rebolo}, {Rich}, {Richards}, {Robin}, {Rocha-Pinto},
  {Rockosi}, {Roe}, {Ross}, {Ross}, {Rossi}, {Rubi{\~n}o-Martin}, {Samushia}, {Sanchez Almeida}, {S{\'a}nchez}, {Santiago}, {Sayres}, {Schlegel}, {Schlesinger}, {Schmidt}, {Schneider}, {Schultheis}, {Schwope}, {Sc{\'o}ccola}, {Seljak}, {Sheldon}, {Shen}, {Shu}, {Simmerer}, {Simmons}, {Skibba}, {Skrutskie}, {Slosar}, {Sobreira}, {Sobeck}, {Stassun}, {Steele}, {Steinmetz}, {Strauss}, {Streblyanska}, {Suzuki}, {Swanson}, {Tal}, {Thakar}, {Thomas}, {Thompson}, {Tinker}, {Tojeiro}, {Tremonti}, {Vargas Maga{\~n}a}, {Verde}, {Viel}, {Vikas}, {Vogt}, {Wake}, {Wang}, {Weaver}, {Weinberg}, {Weiner}, {West}, {White}, {Wilson}, {Wisniewski}, {Wood-Vasey}, {Yanny}, {Y{\`e}che}, {York}, {Zamora}, {Zasowski}, {Zehavi}, {Zhao}, {Zheng}, {Zhu}, \& {Zinn}}]{z}
{Ahn}, C.~P., {Alexandroff}, R., {Allende Prieto}, C., {et~al.} 2012, \apjs, 203, 21

\bibitem[{Akaike(1974)}]{AIC}
Akaike, H. 1974, IEEE Transactions on Automatic Control, 19, 716

\bibitem[{{Alston} {et~al.}(2014){Alston}, {Markevičiūtė., J}, {Kara}, {Fabian}, \& {Middleton}}]{allQPO}
{Alston}, W., {Markevičiūtė., J}, {Kara}, E., {Fabian}, A., \& {Middleton}, M. 2014, MNRAS, 445, L16

\bibitem[{{Arnaud}(1996)}]{xspec}
{Arnaud}, K.~A. 1996, in {Astronomical Society of the Pacific Conference Series}, Vol. 101, {Astronomical Data Analysis Software and Systems V}, ed. G.~H. {Jacoby} \& J.~{Barnes}, 17

\bibitem[{{Bian} \& {Huang}(2010)}]{goodM}
{Bian}, W., \& {Huang}, K. 2010, MNRAS, 401, 507

\bibitem[{{Blandford} \& {Payne}(1982)}]{mag}
{Blandford}, R., \& {Payne}, D. 1982, MNRAS, 199, 883

\bibitem[{{Boller} {et~al.}(1996){Boller}, {Brandt}, \& {Fink}}]{NLS1}
{Boller}, T., {Brandt}, W., \& {Fink}, H. 1996, A\&A, 305, 53

\bibitem[{{Chartas} {et~al.}(2003){Chartas}, {Brandt}, \& {Gallagher}}]{UFO1}
{Chartas}, G., {Brandt}, W.~N., \& {Gallagher}, S.~C. 2003, \apj, 595, 85

\bibitem[{{Chatzikos} {et~al.}(2023){Chatzikos}, {Bianchi}, {Camilloni}, {Chakraborty}, {Gunasekera}, {Guzm{\'a}n}, {Milby}, {Sarkar}, {Shaw}, {van Hoof}, \& {Ferland}}]{cloudyOri}
{Chatzikos}, M., {Bianchi}, S., {Camilloni}, F., {et~al.} 2023, \rmxaa, 59, 327

\bibitem[{{Dauser} {et~al.}(2014){Dauser}, {Garcia}, {Parker}, {Fabian}, \& {Wilms}}]{REL2}
{Dauser}, T., {Garcia}, J., {Parker}, M.~L., {Fabian}, A.~C., \& {Wilms}, J. 2014, MNRAS, 444, L100

\bibitem[{{den Herder} {et~al.}(2001){den Herder}, {Brinkman}, {Kahn}, {Branduardi-Raymont}, {Thomsen}, {Aarts}, {Audard}, {Bixler}, {den Boggende}, {Cottam}, {Decker}, {Dubbeldam}, {Erd}, {Goulooze}, {G{\"u}del}, {Guttridge}, {Hailey}, {Janabi}, {Kaastra}, {de Korte}, {van Leeuwen}, {Mauche}, {McCalden}, {Mewe}, {Naber}, {Paerels}, {Peterson}, {Rasmussen}, {Rees}, {Sakelliou}, {Sako}, {Spodek}, {Stern}, {Tamura}, {Tandy}, {de Vries}, {Welch}, \& {Zehnder}}]{RGS}
{den Herder}, J.~W., {Brinkman}, A., {Kahn}, S.~M., {et~al.} 2001, A\&A, 365, L7

\bibitem[{{Done} {et~al.}(2012){Done}, {Davis}, {Jin}, {Blaes}, \& {Ward}}]{soft2}
{Done}, C., {Davis}, S., {Jin}, C., {Blaes}, O., \& {Ward}, M. 2012, MNRAS, 420, 1848

\bibitem[{{Fabian}(2012)}]{OUTrev1}
{Fabian}, A. 2012, ARA\&A, 50, 455

\bibitem[{{Ferland} {et~al.}(1998){Ferland}, {Korista}, {Verner}, {Ferguson}, {Kingdon}, \& {Verner}}]{cloudy_ori}
{Ferland}, G.~J., {Korista}, K.~T., {Verner}, D.~A., {et~al.} 1998, \pasp, 110, 761

\bibitem[{{Garc{\'\i}a} {et~al.}(2013){Garc{\'\i}a}, {Dauser}, {Reynolds}, {Kallman}, {McClintock}, {Wilms}, \& {Eikmann}}]{XIL3}
{Garc{\'\i}a}, J., {Dauser}, T., {Reynolds}, C.~S., {et~al.} 2013, ApJ, 768, 146

\bibitem[{{Garc{\'\i}a} \& {Kallman}(2010)}]{XIL1}
{Garc{\'\i}a}, J., \& {Kallman}, T.~R. 2010, ApJ, 718, 695

\bibitem[{{Garc{\'\i}a} {et~al.}(2011){Garc{\'\i}a}, {Kallman}, \& {Mushotzky}}]{XIL2}
{Garc{\'\i}a}, J., {Kallman}, T.~R., \& {Mushotzky}, R.~F. 2011, ApJ, 731, 131

\bibitem[{{Garc{\'\i}a} {et~al.}(2014){Garc{\'\i}a}, {Dauser}, {Lohfink}, {Kallman}, {Steiner}, {McClintock}, {Brenneman}, {Wilms}, {Eikmann}, {Reynolds}, \& {Tombesi}}]{REL1}
{Garc{\'\i}a}, J., {Dauser}, T., {Lohfink}, A., {et~al.} 2014, ApJ, 782, 76

\bibitem[{{Gierli{\'n}ski} {et~al.}(2008){Gierli{\'n}ski}, {Middleton}, {Ward}, \& {Done}}]{qpodisc}
{Gierli{\'n}ski}, M., {Middleton}, M., {Ward}, M., \& {Done}, C. 2008, \nat, 455, 369, \dodoi{10.1038/nature07277}

\bibitem[{{Gofford} {et~al.}(2013){Gofford}, {Reeves}, {Tombesi}, {Braito}, {Turner}, {Miller}, \& {Cappi}}]{UFOfrac1}
{Gofford}, J., {Reeves}, J., {Tombesi}, F., {et~al.} 2013, MNRAS, 430, 60

\bibitem[{{Gofford} {et~al.}(2015){Gofford}, {Reeves}, {McLaughlin}, {Braito}, {Turner}, {Tombesi}, \& {Cappi}}]{UFOdrive1}
{Gofford}, J., {Reeves}, J.~N., {McLaughlin}, D.~E., {et~al.} 2015, \mnras, 451, 4169

\bibitem[{{Grant} {et~al.}(2022){Grant}, {Miller}, {Bautz}, {Foster}, {Kraft}, {Allen}, \& {Burrows}}]{HERR}
{Grant}, C.~E., {Miller}, E., {Bautz}, M., {et~al.} 2022, Proc. SPIE Conf., 12181, 121812E

\bibitem[{{Grevesse} \& {Sauval}(1998)}]{AFe}
{Grevesse}, N., \& {Sauval}, A.~J. 1998, \ssr, 85, 161

\bibitem[{{Gunasekera} {et~al.}(2023){Gunasekera}, {van Hoof}, {Chatzikos}, \& {Ferland}}]{CLOUDY}
{Gunasekera}, C., {van Hoof}, P., {Chatzikos}, M., \& {Ferland}, G. 2023, RNAAS, 7, 246

\bibitem[{{HI4PI Collaboration} {et~al.}(2016){HI4PI Collaboration}, {Ben Bekhti}, {Fl{\"o}er}, {Keller}, {Kerp}, {Lenz}, {Winkel}, {Bailin}, {Calabretta}, {Dedes}, {Ford}, {Gibson}, {Haud}, {Janowiecki}, {Kalberla}, {Lockman}, {McClure-Griffiths}, {Murphy}, {Nakanishi}, {Pisano}, \& {Staveley-Smith}}]{hi4pi}
{HI4PI Collaboration}, {Ben Bekhti}, N., {Fl{\"o}er}, L., {et~al.} 2016, \aap, 594, A116

\bibitem[{{Humphrey} {et~al.}(2009){Humphrey}, {Liu}, \& {Buote}}]{cstat}
{Humphrey}, P., {Liu}, W., \& {Buote}, D. 2009, ApJ, 693, 822

\bibitem[{{Igo} {et~al.}(2020){Igo}, {Parker}, {Matzeu}, {Alston}, {Alvarez Crespo}, {Fürst}, {Buisson}, {Lobban}, {Joyce}, {Mallick}, {Schartel}, \& {Santos-Lleó}}]{varufo}
{Igo}, Z., {Parker}, M.~L., {Matzeu}, G.~A., {et~al.} 2020, \mnras, 493, 1088

\bibitem[{{Jansen} {et~al.}(2001){Jansen}, {Lumb}, {Altieri}, {Clavel}, {Ehle}, {Erd}, {Gabriel}, {Guainazzi}, {Gondoin}, {Much}, {Munoz}, {Santos}, {Schartel}, {Texier}, \& {Vacanti}}]{XMM}
{Jansen}, F., {Lumb}, D., {Altieri}, B., {et~al.} 2001, A\&A, 365, L1

\bibitem[{{Jin} {et~al.}(2020){Jin}, {Done}, \& {Ward}}]{wa2}
{Jin}, C., {Done}, C., \& {Ward}, M. 2020, \mnras, 500, 2475

\bibitem[{{Kaastra} \& {Bleeker}(2016)}]{kaastra_bleeker}
{Kaastra}, J., \& {Bleeker}, J. 2016, A\&A, 587, A151

\bibitem[{{Kallman} \& {Bautista}(2001)}]{Lion}
{Kallman}, T., \& {Bautista}, M. 2001, ApJS, 133, 221

\bibitem[{{King} \& {Pounds}(2015)}]{OUTrev2}
{King}, A., \& {Pounds}, K. 2015, ARA\&A, 53, 115

\bibitem[{{Kraemer} {et~al.}(2018){Kraemer}, {Tombesi}, \& {Bottorff}}]{UFOdrive2}
{Kraemer}, S.~B., {Tombesi}, F., \& {Bottorff}, M.~C. 2018, \apj, 852, 35

\bibitem[{{Laha} {et~al.}(2021){Laha}, {Reynolds}, {Reeves}, {Kriss}, {Guainazzi}, {Smith}, {Veilleux}, \& {Proga}}]{OUTrev3}
{Laha}, S., {Reynolds}, C., {Reeves}, J., {et~al.} 2021, Nat. Astron, 5, 13

\bibitem[{{Luminari} {et~al.}(2021){Luminari}, {Nicastro}, {Elvis}, {Piconcelli}, {Tombesi}, {Zappacosta}, \& {Fiore}}]{acc}
{Luminari}, A., {Nicastro}, F., {Elvis}, M., {et~al.} 2021, A\&A, 646, A111

\bibitem[{{Maitra} \& {Miller}(2010)}]{wa4}
{Maitra}, D., \& {Miller}, J.~M. 2010, \apj, 718, 551

\bibitem[{{Mas-Ribas}(2019)}]{collout1}
{Mas-Ribas}, L. 2019, \apj, 885, 95

\bibitem[{{McKinney} {et~al.}(2012){McKinney}, {Tchekhovskoy}, \& {Blandford}}]{MCAF}
{McKinney}, J.~C., {Tchekhovskoy}, A., \& {Blandford}, R.~D. 2012, \mnras, 423, 3083

\bibitem[{{Middleton} \& {Done}(2010)}]{newLedd}
{Middleton}, M., \& {Done}, C. 2010, MNRAS, 403, 9

\bibitem[{{Middleton} {et~al.}(2009){Middleton}, {Done}, {Ward}, {Gierliński}, \& {Schurch}}]{soft}
{Middleton}, M., {Done}, C., {Ward}, M., {Gierliński}, M., \& {Schurch}, N. 2009, MNRAS, 394, 250

\bibitem[{{Middleton} {et~al.}(2011){Middleton}, {Uttley}, \& {Done}}]{wa1}
{Middleton}, M., {Uttley}, P., \& {Done}, C. 2011, \mnras, 417, 250

\bibitem[{{Nardini} {et~al.}(2015){Nardini}, {Reeves}, {Gofford}, {Harrison}, {Risaliti}, {Braito}, {Costa}, {Matzeu}, {Walton}, {Behar}, {Boggs}, {Christensen}, {Craig}, {Hailey}, {Matt}, {Miller}, {O’Brien}, {Stern}, {Turner}, \& {Ward}}]{UFOobs}
{Nardini}, E., {Reeves}, J., {Gofford}, J., {et~al.} 2015, Sci, 347, 860

\bibitem[{{Ogorzalek} {et~al.}(2022){Ogorzalek}, {King}, {Allen}, {Raymond}, \& {Wilkins}}]{collout2}
{Ogorzalek}, A., {King}, A.~L., {Allen}, S.~W., {Raymond}, J.~C., \& {Wilkins}, D.~R. 2022, \mnras, 516, 5027

\bibitem[{{Pounds} {et~al.}(2003){Pounds}, {King}, {Page}, \& {O'Brien}}]{UFO2}
{Pounds}, K.~A., {King}, A.~R., {Page}, K.~L., \& {O'Brien}, P.~T. 2003, \mnras, 346, 1025

\bibitem[{{Proga} \& {Kallman}(2004)}]{radsim}
{Proga}, D., \& {Kallman}, T. 2004, ApJ, 616, 688

\bibitem[{{Proga} {et~al.}(2000){Proga}, {Stone}, \& {Kallman}}]{windsim}
{Proga}, D., {Stone}, J.~M., \& {Kallman}, T.~R. 2000, \apj, 543, 686

\bibitem[{{Puchnarewicz} {et~al.}(1998){Puchnarewicz}, {Mason}, \& {Siemiginowska}}]{BLR}
{Puchnarewicz}, E., {Mason}, K., \& {Siemiginowska}, A. 1998, MNRAS, 293, L52

\bibitem[{{Reeves} {et~al.}(2003){Reeves}, {O’Brien}, \& {Ward}}]{UFO3}
{Reeves}, J.~N., {O’Brien}, P.~T., \& {Ward}, M.~J. 2003, \apj, 593, L65

\bibitem[{{Ricci} {et~al.}(2017){Ricci}, {Trakhtenbrot}, {Koss}, {Ueda}, {Schawinski}, {Oh}, {Lamperti}, {Mushotzky}, {Treister}, {Ho}, {Weigel}, {Bauer}, {Paltani}, {Fabian}, {Xie}, \& {Gehrels}}]{BASS}
{Ricci}, C., {Trakhtenbrot}, B., {Koss}, M., {et~al.} 2017, Nature, 549, 488

\bibitem[{{Shakura} \& {Sunyaev}(1973)}]{disk1}
{Shakura}, N., \& {Sunyaev}, R. 1973, A\&A, 24, 337

\bibitem[{{Smith} \& {Hughes}(2010)}]{tCIE}
{Smith}, R., \& {Hughes}, J. 2010, ApJ, 718, 583

\bibitem[{{Str{\"u}der} {et~al.}(2001){Str{\"u}der}, {Briel}, {Dennerl}, {Hartmann}, {Kendziorra}, {Meidinger}, {Pfeffermann}, {Reppin}, {Aschenbach}, {Bornemann}, {Br{\"a}uninger}, {Burkert}, {Elender}, {Freyberg}, {Haberl}, {Hartner}, {Heuschmann}, {Hippmann}, {Kastelic}, {Kemmer}, {Kettenring}, {Kink}, {Krause}, {M{\"u}ller}, {Oppitz}, {Pietsch}, {Popp}, {Predehl}, {Read}, {Stephan}, {St{\"o}tter}, {Tr{\"u}mper}, {Holl}, {Kemmer}, {Soltau}, {St{\"o}tter}, {Weber}, {Weichert}, {von Zanthier}, {Carathanassis}, {Lutz}, {Richter}, {Solc}, {B{\"o}ttcher}, {Kuster}, {Staubert}, {Abbey}, {Holland}, {Turner}, {Balasini}, {Bignami}, {La Palombara}, {Villa}, {Buttler}, {Gianini}, {Lain{\'e}}, {Lumb}, \& {Dhez}}]{PN}
{Str{\"u}der}, L., {Briel}, U., {Dennerl}, K., {et~al.} 2001, A\&A, 365, L18

\bibitem[{{Tarter} {et~al.}(1969){Tarter}, {Tucker}, \& {Salpeter}}]{xi}
{Tarter}, C., {Tucker}, W., \& {Salpeter}, E. 1969, ApJ, 156, 943

\bibitem[{{Taylor} \& {Reynolds}(2018)}]{disk2}
{Taylor}, C., \& {Reynolds}, C. 2018, ApJ, 868, 109

\bibitem[{{Tombesi} {et~al.}(2010){Tombesi}, {Cappi}, {Reeves}, {Palumbo}, {Yaqoob}, {Braito}, \& {Dadina}}]{UFOsearch1}
{Tombesi}, F., {Cappi}, M., {Reeves}, J.~N., {et~al.} 2010, A\&A, 521, A57

\bibitem[{{Xu} {et~al.}(2024){Xu}, {Pinto}, {Rogantini}, {Barret}, {Bianchi}, {Guainazzi}, {Ebrero}, {Alston}, {Kara}, \& {Cusumano}}]{waufo}
{Xu}, Y., {Pinto}, C., {Rogantini}, D., {et~al.} 2024, A\&A

\bibitem[{{Zhou} {et~al.}(2024){Zhou}, {Mao}, {Fang}, {Wang}, {Nicastro}, \& {Chen}}]{wa3}
{Zhou}, Z., {Mao}, J., {Fang}, T., {et~al.} 2024, \apj, 967, 105

\end{thebibliography}
\bibliographystyle{aasjournal}

\end{document}